\documentclass[a4paper,11pt]{article}

\usepackage{jcappub} 
\usepackage[utf8]{inputenc}
\DeclareUnicodeCharacter{2013}{--}
\DeclareUnicodeCharacter{2014}{---}
\DeclareUnicodeCharacter{2018}{`}
\DeclareUnicodeCharacter{2019}{'}
\DeclareUnicodeCharacter{00E9}{\'e}
\DeclareUnicodeCharacter{00F3}{\'o}
\DeclareUnicodeCharacter{00EA}{\^e}
\DeclareUnicodeCharacter{00E3}{\~a}
\DeclareUnicodeCharacter{00ED}{\'{\i}}
\DeclareUnicodeCharacter{03B1}{$\alpha$}
\DeclareUnicodeCharacter{03A5}{$\Upsilon$}
\DeclareUnicodeCharacter{03B6}{$\zeta$}
\DeclareUnicodeCharacter{03A9}{$\Omega$}
\DeclareUnicodeCharacter{03C1}{$\rho$}
\DeclareUnicodeCharacter{064B}{}
\usepackage[T1]{fontenc} 
\usepackage[utf8]{inputenc}
\newcommand{\ns}{n_\mathrm{s}}
\usepackage{graphicx}
\usepackage{dcolumn}
\usepackage{bm}
\hypersetup{colorlinks,linkcolor=magenta,citecolor=magenta,urlcolor=magenta}
\usepackage{epstopdf,epsf,latexsym,amssymb,amsmath,color,slashed,mathrsfs,mathcomp}
\usepackage{placeins}
\usepackage[center]{subfigure}
\usepackage{multirow}
\usepackage{makecell}
\usepackage[table]{xcolor}
\usepackage{booktabs}
\newcommand{\orcid}[1]{\href{https://orcid.org/#1}{\includegraphics[width=10pt]{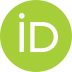}}}

\graphicspath{{figs/}}

\title{\textbf{Constraining Quintessential Inflation with ACT: A Gauss-Bonnet Gateway}}
\author[a,d]{Yogesh\orcid{0000-0002-7638-3082},}
\author[b]{Imtiyaz Ahmad Bhat\orcid{0000-0002-2695-9709},}
\author[c]{Mayukh R. Gangopadhyay\orcid{0000-0002-1466-8525},}
\author[d,e]{M. Sami\orcid{0000-0003-1081-0632}}
\affiliation[a]{Institute for Theoretical Physics and Cosmology,
Zhejiang University of Technology, Hangzhou 310023, China}
\affiliation[b]{Department of Physics, Islamic University of Science and Technology, Awantipora, J\&K-192122, India}
\affiliation[c]{Department of Physics, School of Advanced Sciences, Vellore Institute of Technology (VIT),
Chennai Campus, Chennai 600127, Tamil Nadu, India}
\affiliation[d]{Centre for Cosmology and Science Popularization (CCSP), SGT University, Gurugram, Haryana, 122505, India}
\affiliation[e]{Eurasian International Centre for Theoretical Physics, Astana, Kazakhstan}
\emailAdd{yogesh@zjut.edu.cn, yogeshjjmi@gmail.com}
\emailAdd{bhat.imtiyaz@iust.ac.in}
\emailAdd{mayukhraj.g@vit.ac.in}
\emailAdd{sami\_ccsp@sgtuniversity.org}
\abstract{
Recent results from the Atacama Cosmology Telescope (ACT), which, when combined with Planck, CMB lensing, DESI DR2 with BAO data, indicate a higher and more tightly constrained scalar spectral index, $\ns = 0.9743 \pm 0.0034$ P-ACT-LB), place several inflationary models under tension, with quintessential inflation pushed close to or beyond the $2\sigma$ boundary in the $r$--$\ns$ plane for this dataset combination. In this work, we revisit quintessential inflation within the framework of Einstein--Gauss--Bonnet (EGB) gravity, where a scalar field non-minimally coupled to the Gauss--Bonnet invariant modifies the inflationary dynamics. We consider three representative coupling functions -- exponential, hyperbolic secant, and hyperbolic tangent -- and show that the exponential and sech-type couplings can shift the predicted values of $r$ and $\ns$ into the $1\sigma$ region allowed by the P-ACT-LB-BK18 combination, thereby restoring consistency with these observations. In contrast, the tanh-type coupling remains disfavoured, underscoring the sensitivity of inflationary observables to the coupling structure. We further investigate the reheating phase using a model-independent parameterisation and demonstrate that viable thermal histories can be realised even in the absence of a potential minimum, with reheating temperatures consistent with Big Bang nucleosynthesis bounds. We emphasise that the quantitative conclusions are conditional on the choice of dataset combination: the preference for a higher $\ns$ is driven by the ACT data and, indirectly, by DESI BAO, whereas combinations based on Planck alone or including the latest SPT-3G data yield lower central values of $\ns$ and correspondingly milder tension for standard quintessential inflation. Overall, our analysis shows that EGB corrections provide a viable and robust extension that can reconcile quintessential inflation with the ACT-based precision cosmological data, and we identify the corresponding allowed parameter space.
}
\begin{document}
\maketitle
\section{\label{introduction}Introduction}
Inflation remains the leading paradigm for the physics of the very early universe. Originally proposed to resolve the flatness, horizon, and monopole problems of standard Big Bang cosmology~\cite{Guth:1980zm,Sato:1981qmu,Linde:1981mu,PhysRevLett.48.1220}, it also provides a natural mechanism for generating the primordial density perturbations that seed large-scale structure~\cite{Mukhanov:1981xt,starobinsky:1982ee,starobinsky:1983zz,Starobinsky:1980te}. These quantum fluctuations, amplified and stretched to cosmological scales during the quasi-de Sitter expansion, have been measured with increasing precision over the past two decades~\cite{WMAP:2012nax,Planck:2015sxf,Planck:2018jri}. Despite this observational success, the precise particle physics realization of inflation remains an open question, and the final test of the inflationary paradigm is yet to come.

Inflationary model building has taken many forms, from simple single-field slow-roll scenarios to more elaborate constructions involving non-canonical kinetic terms, non-minimal couplings, and modifications of gravity~\cite{Barenboim:2007ii,Odintsov:2025eiv,Franche:2010yj,Unnikrishnan:2012zu,Saaidi:2015kaa,Mohammadi:2015jka,Fairbairn:2002yp,Aghamohammadi:2014aca,Bessada:2009pe,Weller:2011ey,Nazavari:2016yaa,Amani:2018ueu,Mohammadi:2018zkf,berera1995warm,berera2000warm,BasteroGil:2004tg, Bastero-Gil:2017wwl, Gangopadhyay:2020bxn, Gangopadhyay:2026mck, Basak:2021cgk, Gangopadhyay:2026vdu, Gangopadhyay:2026xqj, Sayar:2017pam,Akhtari:2017mxc,Sheikhahmadi:2019gzs,Motohashi:2014ppa,Odintsov:2017hbk,Oikonomou:2017bjx,Mohammadi:2019qeu,Mohammadi:2020ftb,Mohammadi:2022vru,Adhikari:2020xcg,Yogesh:2024zwi,Yang:2024tkw}. Modified gravity theories have proven particularly fruitful in this regard, since higher-curvature corrections can dramatically reshape the inflationary dynamics and revive potentials that would otherwise be in conflict with observations within standard Einstein gravity~\cite{Mohammadi:2023kzd,Sahni:2001qp,Barvinsky:1994hx,Cervantes-Cota:1995ehs,Barvinsky:2008ia,DeSimone:2008ei,Gialamas:2020vto,Bezrukov:2008ej,Kallosh:2013pby,Barvinsky:2009ii,Rubio:2018ogq,Gialamas:2023flv,Gialamas:2022xtt,Gialamas:2021enw,Gialamas:2020snr,Kim:2025ikw,Kim:2025dyi,Koshelev:2020xby}.\footnote{Inflation is expected to occur at energy scales well above the electroweak scale, up to near the grand unified theory (GUT) scale. At lower energy scales, much of the interesting beyond-the-Standard-Model physics -- including the generation of the baryon asymmetry, dark matter production, and phase transitions -- complicates the inflationary picture and makes a clean realization of slow-roll inflation more difficult.}

The situation has been sharpened considerably by the recent data release from the Atacama Cosmology Telescope (ACT). The ACT DR6 collaboration has reported $n_s = 0.9709 \pm 0.0038$ from a combined Planck and ACT analysis, which tightens further to $n_s = 0.9743 \pm 0.0034$ upon inclusion of cosmic microwave background (CMB) lensing, baryon acoustic oscillation (BAO), and Dark Energy Spectroscopic Instrument (DESI) DR2 data~\cite{ACT:2025fju,ACT:2025tim,DESI:2025zpo}. This upward shift in the central value of $n_s$, together with the improved precision, places a number of previously viable models under significant strain. In particular, the quintessential inflation model~\cite{Peebles_1999} - which unifies inflation with dark energy through a single scalar field carrying a runaway potential -- was broadly consistent with earlier Planck data but now finds itself near or beyond the $2\sigma$ boundary of the ACT-allowed region. This has prompted a widespread reassessment of inflationary models in light of ACT~\cite{Odintsov:2025wai,Kallosh:2025rni,Aoki:2025wld,Dioguardi:2025vci,Salvio:2025izr,Brahma:2025dio,Gao:2025onc,Drees:2025ngb,Zharov:2025evb,Yin:2025rrs,Liu:2025qca,Gialamas:2025ofz,Haque:2025uri,Haque:2025uis,Dioguardi:2025mpp,Gialamas:2025kef,Yogesh:2025wak,Mohammadi:2025gbu,risdianto2025,ferreira2025,mohammadi2025,pallis2025b,gao2025,okada2025,mcdonald2025,bianchi2025,odintsov2025,kohri2025,chakraborty2025,pallis2025,frolovsky2025,haque2025,peng2025,bernardo2025,yi2025,addazi2025,maity2025,byrnes2025,zharov2025,Mondal:2025kur,Wolf:2025ecy,Ye:2025idn,Zahoor:2025nuq,Oikonomou:2026gae,ZentenoGatica:2026pbf,Ellis:2025zrf,Dorsch:2026fgb,Sabogal:2026qvy,Verdiani:2026cpc,DelGrosso:2026zbg,Qiu:2026npi,Yang:2026rzn,Nashed:2026imt,Talebian:2026fnn,Fomin:2026siz,Nojiri:2026ish,Waeming:2026wrp, Giare:2026oti,Nerval:2026iev,Gonuguntla:2026rkw,DiBenedetto:2026onj,Ahmed:2026gqq,Kallosh:2026qrc,Kaneta:2026qxx,Pirzada:2026sle,Yuan:2026xcg}. The interpretation of the tight constraint $\ns = 0.9743 \pm 0.0034$, by P-ACT-LB-BK18  (Planck CMB anisotropies combined with CMB lensing [L], DESI DR2 BAO [B], and BICEP/Keck 2018 B-mode data [BK18] and ACT) which motivates this study and serves as the observational benchmark of this work. First, the datasets we are considering here individually are mildly differ on $n_s$ within the baseline $\Lambda$CDM$+r$ model: Planck 2018 alone gives $n_s = 0.9649 \pm 0.0042$~\cite{Planck:2018jri}, roughly $1\sigma$ below the joint Planck+ACT (P-ACT) value $\ns = 0.9709 \pm 0.0038$~\cite{ACT:2025fju,ACT:2025tim}. Second, the further upward shift induced by DESI DR2 BAO data is an \emph{indirect} effect: DESI prefers a lower matter density $\Omega_m$ than CMB data alone, and the combined fit responds by linking $\Omega_m$ with $\ns$, pushing the $n_s$ to higher values~\cite{ACT:2025fju,ACT:2025tim,DESI:2025zpo,ferreira2025}. Third, exactly because the individual posteriors are only partially overlapping and combining them leads to formally narrow uncertainties, inflationary model selection implications are therefore combination-dependent. Finally, ACT is not the sole modern-day high-resolution ground-based CMB experiment:  the Third Generation Camera on the South Pole Telescope (SPT-3G) D1 release~\cite{SPT-3G:2025bzu} finds~\footnote{The SPT-3G constraint is represented as a vertical rectangular band because it strongly constrains the $n_s$ but has limited independent sensitivity to the tensor-to-scalar ratio $r$. The rectangle represents the one-dimensional confidence interval on $n_s$, which is stretched to some range of $r$ shown in the Figs.~\ref{r_ns_exp},\ref{r_ns_sech} and Fig~\ref{r_ns_tanh} by the brown shaded region.}, in combination with Planck, $\ns \simeq 0.9636 \pm 0.0035$, consistent with the original Planck value and lower than the P-ACT result, while the combination CMB-SPA-ACT (denoted as CMB-SPA) gives $\ns = 0.9684 \pm 0.0030$, adding the DESI BAO data would yields $\ns \simeq 0.9728 \pm 0.0027$~\cite{SPT-3G:2025bzu,SPT-3G:2025vyw}. A significantly higher $\ns$ is thus not universally present across current CMB datasets. Throughout this paper we adopt the P-ACT-LB(-BK18) combination as our benchmark, in line with much of the recent literature, but the reader should bear in mind that the degree of tension of quintessential inflation and hence the phenomenological pressure motivating the Gauss Bonnet extension is contingent on this choice.

Following the end of inflation, the universe enters a cold, low-entropy state, and a reheating phase is necessary to repopulate it with relativistic particles~\cite{Kofman:1994rk}. During reheating, the inflaton field decays into Standard Model particles, which thermalise the primordial plasma and initiate the radiation-dominated epoch. This process eventually connects inflation to Big Bang nucleosynthesis (BBN); see Refs.~\cite{Lozanov:2019jxc,Kofman:1997pt} for comprehensive reviews. While BBN sets a firm lower bound on the reheating temperature, $T_{\rm BBN} \simeq 4~{\rm MeV}$, the upper bound from inflation itself, $T_{\rm re} \lesssim 10^{16}~{\rm GeV}$, leaves a physically wide window~\cite{Cook:2015vqa,Hasegawa:2019jsa,Kawasaki:1999na,Kawasaki:2000en}. In models with non-minimal couplings, including those studied here, the absence of a potential minimum means that reheating cannot proceed via coherent oscillations, and a generalized treatment is required.

Non-minimally coupled scalar field models, including models with coupling to the Gauss--Bonnet term, have long been known to yield reduced values of the tensor-to-scalar ratio~\cite{Bezrukov:2007ep,Bezrukov:2010jz}, and have attracted renewed interest in the post-ACT landscape where a tighter upper bound $r < 0.038$ (P-ACT-LB-BK18) further constrains model space. From a theoretical perspective, the Gauss--Bonnet term arises naturally as a quantum correction to the Einstein--Hilbert action in the low-energy effective action of string theory, making it a well-motivated ingredient in any description of early-universe physics.

In this paper, we study quintessential inflation in the Einstein--Gauss--Bonnet (EGB) framework. The scalar field is non-minimally coupled to the Gauss-Bonnet invariant through a coupling function $\xi(\phi)$, which modifies the dynamics in a way that can restore agreement with the P-ACT-LB-BK18 observational constraints. We consider three physically motivated coupling functions: an exponential form and two hyperbolic forms (sech and tanh)~\cite{TerenteDiaz:2023iqk,vandeBruck:2015gjd,Guo:2009uk,Koh:2016abf,Satoh:2008ck,Jiang:2013gza,Koh:2018qcy,Mathew:2016anx,Pozdeeva:2020apf,Pozdeeva:2016cja,Nozari:2017rta,Odintsov:2018zhw,Fomin:2019yls,Odintsov:2020sqy,Odintsov:2020zkl,Kawai:2021bye,Kawai:2017kqt,Oikonomou:2022xoq,Oikonomou:2022ksx,Cognola:2006sp,Odintsov:2020xji,Nojiri:2019dwl,Oikonomou:2024etl,Oikonomou:2024jqv,Odintsov:2023weg,Kawai:1998ab,Nojiri:2024hau,Elizalde:2023rds,Nojiri:2023mvi,Odintsov:2023aaw,Odintsov:2022rok,Kawai:2023nqs,Lopez:2025gfu,Pozdeeva:2020shl,Yi:2018gse,Kleidis:2019ywv,Rashidi:2020wwg,Kawai:2021edk,Ashrafzadeh:2023ndt}. We find that the exponential and sech-type couplings can successfully bring the inflationary predictions within the $1\sigma$ region of the P-ACT-LB-BK18 constraints, while the tanh coupling is unable to do so for the parameter ranges explored.

The remainder of the paper is organized as follows. In Sec.~\ref{sec2} we briefly review inflation in EGB gravity and derive the governing equations. In Sec.~\ref{effectivepotential} we introduce the effective potential formalism and express the slow-roll parameters and inflationary observables in its terms. Sec.~\ref{sec:inflation} describes the quintessential inflation potential and the reheating formalism. The results for the exponential coupling are presented in Sec.~\ref{expc}, and those for the hyperbolic couplings in Sec.~\ref{sec:result}. We conclude with a summary in Sec.~\ref{sec:conclusion}.

\section{Inflation in EGB Gravity}
\label{sec2}

We work with a scalar-tensor theory of gravity in which the scalar field $\phi$ is non-minimally coupled to the Gauss--Bonnet invariant. Setting the reduced Planck mass $M_p = 1$, the action takes the form~\cite{Pozdeeva:2020apf,Guo:2009uk}:
\begin{equation}
\label{action1}
S=\int d^4x\sqrt{-g}\left[UR-\frac{g^{\mu\nu}}{2}\partial_\mu\phi\partial_\nu\phi-V(\phi)-\frac{\xi(\phi)}{2}\mathcal{G}\right],
\end{equation}
where $V(\phi)$ and $\xi(\phi)$ are smooth functions of the scalar field, $U$ is a positive constant, and $\mathcal{G}$ is the Gauss--Bonnet invariant:
\begin{equation}
\mathcal{G}=R_{\mu\nu\rho\sigma}R^{\mu\nu\rho\sigma}-4R_{\mu\nu}R^{\mu\nu}+R^2.
\end{equation}

In a spatially flat Friedmann--Lema\^{i}tre--Robertson--Walker (FLRW) universe, varying the action~\eqref{action1} with respect to the metric and the scalar field yields the following equations of motion~\cite{vandeBruck:2015gjd,Pozdeeva:2019agu}:
\begin{eqnarray}
12UH^2 &=& \dot\phi^2+2V+24\dot{\xi}H^3,  \label{Equ0} \\
4U\dot{H} &=& {}-\dot\phi^2+4\ddot{\xi}H^2+4\dot{\xi}H\left(2\dot{H}-H^2\right), \label{EquH}\\
\ddot{\phi}&=&{}-3H\dot{\phi}-V'-12\xi'H^2\left(\dot{H}+H^2\right), \label{Equphi}
\end{eqnarray}
where dots and primes denote derivatives with respect to cosmic time $t$ and the scalar field $\phi$, respectively, and $H = \dot{a}/a$ is the Hubble parameter.

We define the hierarchy of slow-roll parameters following Refs.~\cite{vandeBruck:2015gjd}:
\begin{eqnarray}
\epsilon_1 &=&{}-\frac{\dot{H}}{H^2}={}-\frac{d\ln H}{dN},\qquad \epsilon_{i+1}= \frac{d\ln|\epsilon_i|}{dN},\quad i\geqslant 1, \\
\delta_1&=& \frac{2}{U}H\dot{\xi}=\frac{2}{U}H^2\xi'\frac{d\phi}{dN},\qquad \delta_{i+1}=\frac{d\ln|\delta_i|}{dN},\quad i\geqslant 1,
\end{eqnarray}
where we have used $d/dt = H\,d/dN$. The slow-roll approximation then requires $|\epsilon_i| \ll 1$ and $|\delta_i| \ll 1$.

Imposing these conditions on Eqs.~\eqref{Equ0}--\eqref{Equphi}, we first obtain
\begin{eqnarray}
&&12UH^2 \simeq \dot\phi^2+2V,  \label{Equ0slr1} \\
&&4U\dot{H} \simeq{}-\dot\phi^2-4\dot{\xi}H^3={}-\dot{\phi}\left(\dot{\phi}+4\xi'H^3\right). \label{EquHslr1}
\end{eqnarray}
Using the approximate expression for $\epsilon_1$,
\begin{equation*}
\epsilon_1\simeq \frac{\dot\phi^2}{3(\dot\phi^2+2V)}+\frac{1}{2}\delta_1\ll 1,
\end{equation*}
one finds $\dot\phi^2 \ll 2V$, and Eq.~\eqref{Equ0slr1} reduces to the familiar relation
\begin{equation}
\label{Equ0slr2}
6UH^2 \simeq V.
\end{equation}
Differentiating Eq.~\eqref{Equ0slr2} with respect to time and using Eq.~\eqref{EquHslr1}, we obtain the slow-roll equation for the field velocity:
\begin{equation}
\label{Equphislr1}
\dot{\phi}\simeq{}-\frac{V'}{3H}-4\xi'H^3.
\end{equation}
Substituting this back into Eq.~\eqref{Equphi} confirms $|\ddot{\phi}|\ll|12\xi'H^4|$, consistent with the slow-roll assumption.

Collecting the leading-order slow-roll equations, we have:
\begin{eqnarray}
H^2&\simeq&\frac{V}{6U}\,, \label{Equ0lo}\\
\dot{H}&\simeq&{}-\frac{\dot\phi^2}{4U}-\frac{\dot{\xi}H^3}{U}\,, \label{EquHlo}\\
\dot{\phi}&\simeq&{}-\frac{V'+12\xi'H^4}{3H}. \label{Equphilo}
\end{eqnarray}
These equations form the basis of our inflationary analysis, which we cast in terms of the effective potential introduced in the next section.

\section{The Effective Potential Formalism}
\label{effectivepotential}

\subsection*{Slow-roll parameters in terms of $V_{\rm eff}$}

A particularly elegant way to analyze inflationary dynamics in EGB gravity is through the effective potential~\cite{Pozdeeva:2019agu}:
\begin{equation}
\label{Veff}
V_{\rm eff}(\phi)={}-\frac{U^2}{V(\phi)}+\frac{1}{3}\xi(\phi).
\end{equation}
This effective potential completely characterizes the existence and stability of de Sitter solutions in the EGB framework. For $V(\phi) > 0$, which we assume throughout, it is well defined and provides a compact organizing principle for the inflationary phase.

Using Eqs.~\eqref{EquHlo} and~\eqref{Equphilo}, the leading-order equations for $H$ and $\phi$ as functions of the number of e-folds $N$ become:
\begin{eqnarray}
\frac{dH}{dN}&\simeq&{}-\frac{H}{U}V'V_{\rm eff}'\,, \label{EquHloN}\\
\frac{d\phi}{dN}&\simeq&{}-2\frac{V}{U}V_{\rm eff}'. \label{EquphiloN}
\end{eqnarray}

In terms of the effective potential, the slow-roll parameters read:
\begin{equation}
\epsilon_1 = \frac{V'}{U}V_{\rm eff}'\,,
\label{EGBsr1}
\end{equation}
\begin{equation}
\epsilon_2 = {}-\frac{2V}{U}V_{\rm eff}'\left[\frac{V''}{V'}+\frac{V_{\rm eff}''}{V_{\rm eff}'}\right],
\label{EGBsr2}
\end{equation}
\begin{equation}
\delta_1 = {}-\frac{2V^2}{3U^3}\xi'V_{\rm eff}'\,,
\label{EGBsr3}
\end{equation}
\begin{equation}
\delta_2 = {}-\frac{2V}{U}V_{\rm eff}'\left[2\frac{V'}{V}+\frac{V_{\rm eff}''}{V_{\rm eff}'}+\frac{\xi''}{\xi'}\right].
\label{slrVeffd}
\end{equation}

The slow-roll conditions $|\epsilon_1| \ll 1$ and $|\delta_1| \ll 1$ are both satisfied when $V_{\rm eff}'$ is sufficiently small, which is the underlying geometric condition that sustains inflation in the EGB setting.

The tensor-to-scalar ratio and scalar spectral index are given by~\cite{Koh:2016abf,Guo:2009uk}:
\begin{equation}
\label{rVeff}
r = 8|2\epsilon_1-\delta_1|,
\end{equation}
\begin{equation}
\label{nsVeff}
n_s = 1-2\epsilon_1-\frac{2\epsilon_1\epsilon_2-\delta_1\delta_2}{2\epsilon_1-\delta_1}.
\end{equation}
The amplitude of scalar perturbations at leading order in slow-roll is~\cite{vandeBruck:2015gjd}:
\begin{equation}
\label{As}
A_s\approx\frac{H^2}{\pi^2 U r}\approx\frac{V}{6\pi^2 U^2 r}.
\end{equation}

It is well-known that Einstein-Gauss-Bonnet inflation, considered in the paper, belongs to a general class of models known as Horndeski gravity.
Action~\ref{action1} is a specific point in the space of Horndeski (equivalently,
generalised Galileon) theories, the most general four-dimensional scalar-tensor theories with second-order field equations, originally
classified by Horndeski and shown to coincide with the generalised Galileons by Kobayashi, Yamaguchi and Yokoyama~\cite{Kobayashi2011}. Thus, a consistency check is something on the cards.
In the Appendix of that work, it is shown that a term $f(\phi)\mathcal G$ added to the action is reproduced, up to a total derivative, by the
generalised-Galileon functions
\begin{equation}
G_2=8f^{(4)}X^2(3-\ln X),\quad G_3=4f^{(3)}X(7-3\ln X),
\end{equation}
\begin{equation}
G_4=4f^{(2)}X(2-\ln X),\quad G_5=-4f^{(1)}\ln X,
\end{equation}
where $f^{(n)}\equiv d^nf/d\phi^n$ and
$X\equiv-\tfrac12g^{\mu\nu}\partial_\mu\phi\partial_\nu\phi$. Our coupling
term is $-\xi(\phi)\mathcal G/2$, i.e.\ $f=-\xi/2$, so together with the
canonical kinetic term, the potential $V(\phi)$, and the constant piece
$U$ entering $G_4$ through $UR$, action (\ref{action1}) corresponds to the Horndeski
point
\begin{subequations}\label{eq:horndeski-dictionary}
\begin{align}
G_2(\phi,X) &= X - V(\phi) - 4\xi^{(4)}(\phi)\,X^2(3-\ln X),\\
G_3(\phi,X) &= -2\xi'''(\phi)\,X(7-3\ln X),\\
G_4(\phi,X) &= U - 2\xi''(\phi)\,X(2-\ln X),\\
G_5(\phi,X) &= 2\xi'(\phi)\ln X.
\end{align}
\end{subequations}
In the notation of Ref.~\cite{Kobayashi2011}, the tensor quadratic action
built from these functions is $S_T^{(2)}=\tfrac18\int dt\,d^3x\,a^3
\big[\mathcal G_T\dot h_{ij}^2-a^{-2}\mathcal F_T(\vec\nabla h_{ij})^2\big]$,
with
\begin{equation}
\mathcal F_T = 2\big[G_4-X(\ddot\phi\,G_{5X}+G_{5\phi})\big],\qquad
\mathcal G_T = 2\big[G_4-2XG_{4X}-X(H\dot\phi\,G_{5X}-G_{5\phi})\big],
\end{equation}
and an analogous pair $(\mathcal F_S,\mathcal G_S)$, built from
$G_2,\ldots,G_5$ and the background quantities of
Ref.~\cite{Kobayashi2011}, fixes the scalar sector; the two power spectra
follow as $\mathcal P_T\propto H^2/(\mathcal G_Tc_T^3)$ and
$\mathcal P_\zeta\propto H^2/(\mathcal G_Sc_S^3)$ evaluated at sound-horizon crossing, with $c_T^2=\mathcal F_T/\mathcal G_T$ and
$c_S^2=\mathcal F_S/\mathcal G_S$. Because $(\mathcal F_T,\mathcal G_T,
\mathcal F_S,\mathcal G_S)$ are completely fixed once the functions in Eq.~\eqref{eq:horndeski-dictionary} and the background are specified,
their slow-roll expansion is not an independent calculation, but the same one already carried out via the covariant ADM (Hwang--Noh) approach
rather than the $(\mathcal F,\mathcal G)$ language above in Refs.~\cite{vandeBruck:2015gjd, Guo:2009uk, Guo:2010jr, Yi:2018gse},
which is where Eqs.~(\ref{EGBsr1})--(\ref{As}) originate. Concretely, Ref.~\cite{Yi:2018gse}, which explicitly identifies the Gauss-Bonnet
coupling as a Horndeski subclass, independently derives this same action,
\begin{equation}
n_s-1=-2\epsilon_1-\frac{2\epsilon_1\epsilon_2-\delta_1\delta_2}{2\epsilon_1-\delta_1},
\qquad r=16\epsilon_1-8\delta_1,
\end{equation}
which coincides term by term with our Eqs.~(\ref{nsVeff}) and (\ref{rVeff}) (the latter up to
the absolute value we introduce to keep $r\ge0$ once $\delta_1$ is allowed to change sign, as it does for the tanh coupling of Sec.~\ref{sec:tanh_failure}).
This term-by-term agreement between two independently developed
formalisms: the Horndeski quadratic-action route of
Ref.~\cite{Kobayashi2011} and the covariant ADM route of
Refs.~\cite{vandeBruck:2015gjd, Guo:2009uk, Guo:2010jr, Yi:2018gse}
which confirms that Eqs.~(\ref{EGBsr1})--(\ref{As}) are the correct slow-roll limit of the
general Horndeski/generalised-Galileon perturbation spectra.
\section{The Inflation Model and Reheating}
\label{sec:inflation}

\subsection{Quintessential Inflation in EGB Gravity}

We take the scalar field potential to be of the quintessential inflation form:
\begin{equation}
V(\phi)= V_0\, {\rm e}^{-\lambda \phi^n},
\end{equation}
which serves the dual purpose of driving inflation at early times and dark energy at late times. While this potential is disfavored in standard general relativity by the recent ACT observations, we will show that coupling it to the Gauss--Bonnet term can restore its viability.

For the coupling function, we consider three distinct forms:
\begin{equation}
\xi(\phi) = \frac{1}{V_0}{\rm e}^{-\xi_1 \phi}, \quad
\xi(\phi) = \frac{1}{V_0}{\rm sech}(\xi_1 \phi), \quad
\xi(\phi) = \frac{1}{V_0}{\rm tanh}(\xi_1 \phi),
\label{coupling_function}
\end{equation}
where $\xi_1$ is a free coupling parameter. The overall factor of $V_0^{-1}$ is introduced to simplify the algebra and does not affect the generality of the analysis.

\subsection{Reheating after Quintessential Inflation}

Following the end of inflation, the universe enters a cold, dilute state. Since the quintessential potential has no minimum, the standard picture of coherent oscillation and perturbative decay is not applicable here. Instead, we adopt the generalized reheating formalism, in which the reheating phase is parametrized by an effective equation-of-state parameter $w_{\rm re}$, assumed constant during reheating~\cite{Cook:2015vqa,Bhat:2020wdc,Antoniadis:2026bys}.

Working indirectly through the inflationary observables, one can extract the number of e-folds and temperature of reheating without specifying the details of the decay mechanism~\cite{Cook:2015vqa,Ballardini:2024ado,Gialamas:2019nly}:
\begin{equation}
N_{\rm re}= \frac{4}{1-3w_{\rm re}}\left[61.488 - \ln\!\left(\frac{V_{\rm end}^{1/4}}{H_k}\right) - N_k\right],
\label{Nre}
\end{equation}
\begin{equation}
T_{\rm re}= \left[\left(\frac{43}{11 g_{\rm re}}\right)^{1/3}\frac{a_0 T_0}{k}H_k e^{-N_k}\left(\frac{45\, V_{\rm end}}{\pi^2 g_{\rm re}}\right)^{\!\beta}\right]^{\!\gamma},
\label{Tre}
\end{equation}
where $N_{\rm re}$ and $T_{\rm re}$ are the number of reheating e-folds and the reheating temperature, respectively. The parameter $N_k$ denotes the number of e-folds from horizon crossing to the end of inflation, $g_{\rm re}$ is the number of relativistic degrees of freedom at reheating, and the constants are defined as $\beta = -1/[3(1+w_{\rm re})]$ and $\gamma = \beta/(1-3w_{\rm re})$.

\section{Exponential Coupling}
\label{expc}

\begin{figure*}[t]
    \centering
    \subfigure[\label{r_ns_exp_n}]{%
        \includegraphics[width=0.45\textwidth]{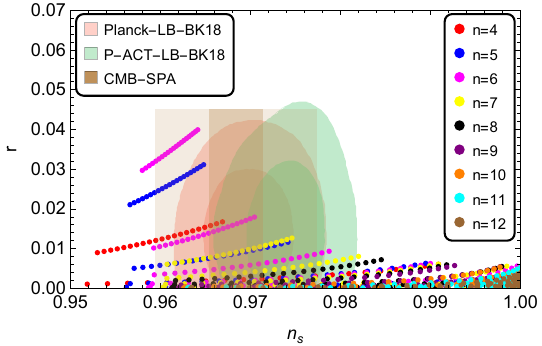}
    }\hfill
    \subfigure[\label{r_ns_exp_lam}]{%
        \includegraphics[width=0.45\textwidth]{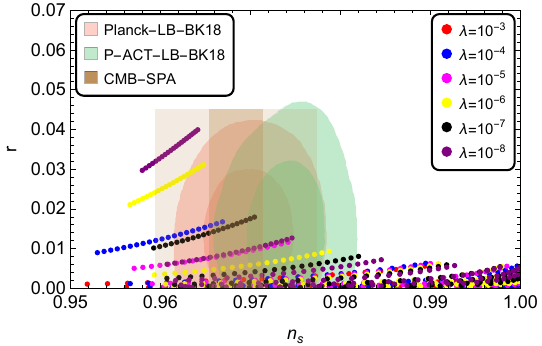}
    }
\caption{The $r$--$n_s$ plane for the exponential Gauss-Bonnet coupling. Left: trajectories for different values of $n$, with $\lambda \in [10^{-8}, 10^{-3}]$ and $\xi_1 \in [0, 0.03]$ varied. Right: trajectories for different values of $\lambda$, with $n \in [4, 12]$ and $\xi_1 \in [0, 0.03]$ varied.
The background dark and light green contours show the P-ACT-LB-BK18 (Planck CMB anisotropies, ACT, combined with CMB lensing
[L], DESI DR2 BAO [B], and BICEP/Keck 2018 B-mode data [BK18]) $1\sigma$ and $2\sigma$ constraint regions. Whereas the dark and light pink shaded regions denote the $1\sigma$ and $2\sigma$ bounds of Planck-LB-BK18. The $1\sigma$ and $2\sigma$ bounds of the Planck-SPT-ACT (denoted as CMB-SPA) are represented by the dark and light brown rectangles, respectively~\cite{SPT-3G:2025bzu,Ellis:2025zrf}.}
\label{r_ns_exp}
\end{figure*}

We begin with the exponential coupling $\xi(\phi) = V_0^{-1} e^{-\xi_1\phi}$. To evaluate the inflationary observables, we compute the slow-roll parameters from Eqs.~\eqref{EGBsr1}--\eqref{slrVeffd}, and solve the field evolution equation~\eqref{EquphiloN} numerically with the end of inflation at $N = 60$ and $N_\star = 0$ corresponding to horizon crossing. The initial condition for $\phi$ is fixed by setting $\epsilon_1 = 1$ at the end of inflation. We explore the parameter ranges $n = 4$--$12$, $\lambda = 10^{-8}$--$10^{-3}$, and $\xi_1 = 0$--$0.02$, choosing $\xi_1$ values that keep the predictions consistent with ACT.

The resulting $r$--$n_s$ trajectories are shown in Fig.~\ref{r_ns_exp}. In the left panel (Fig.~\ref{r_ns_exp_n}), we plot 12 representative values of $n$ while varying $\lambda$ and $\xi_1$; in the right panel (Fig.~\ref{r_ns_exp_lam}), we fix several values of $\lambda$ and vary $n$ and $\xi_1$. Each plotted point corresponds to a distinct combination of $(\lambda, n, \xi_1)$.

Two trends are visible. First, for a fixed coupling $\xi_1$, increasing $n$ tends to reduce the tensor-to-scalar ratio $r$, as seen in the left panel. Second, both panels confirm that for appropriate parameter combinations, the model predictions can be brought inside the $1\sigma$ region of the ACT constraints. It is worth emphasizing that the specific choice of $\xi_1$ is not uniquely determined by theory; rather, the results presented here represent a viable slice of a larger parameter space.

\section{Hyperbolic Couplings}
\label{sec:result}

\begin{figure*}[t]
    \centering
    \subfigure[\label{gwN60}]{\includegraphics[width=0.45\linewidth]{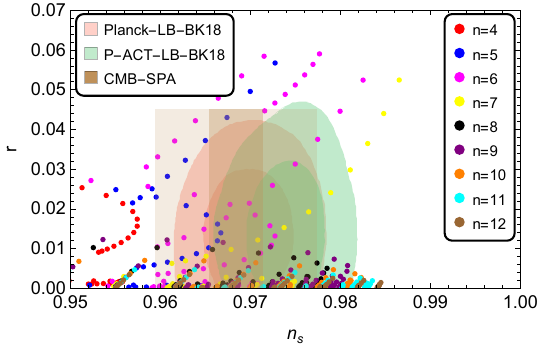}}
    \subfigure[\label{gwN61}]{\includegraphics[width=0.45\linewidth]{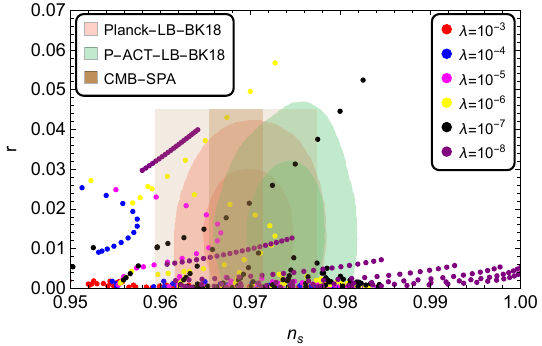}}
    \caption{The $r$--$n_s$ plane for the sech-type Gauss-Bonnet coupling, with $\xi_1 \in [0, 0.2]$. Left: varying $n$. Right: varying $\lambda$. The background dark and light green contours show the P-ACT-LB-BK18 (Planck CMB anisotropies, ACT, combined with CMB lensing
[L], DESI DR2 BAO [B], and BICEP/Keck 2018 B-mode data [BK18]) $1\sigma$ and $2\sigma$ constraint regions. Whereas the dark and light pink shaded regions denote the $1\sigma$ and $2\sigma$ bounds of Planck-LB-BK18. The $1\sigma$ and $2\sigma$ bounds of the Planck-SPT-ACT (denoted as CMB-SPA) are represented by the dark and light brown rectangles, respectively~\cite{SPT-3G:2025bzu,Ellis:2025zrf}.} Both panels show good agreement with the ACT constraints.
    \label{r_ns_sech}
\end{figure*}

\begin{figure*}[t]
    \centering
    \subfigure[]{\includegraphics[width=0.45\linewidth]{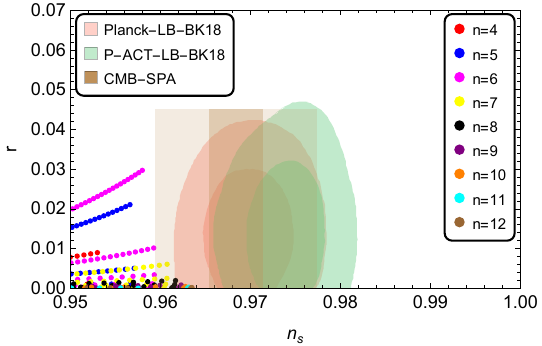}}
    \subfigure[]{\includegraphics[width=0.45\linewidth]{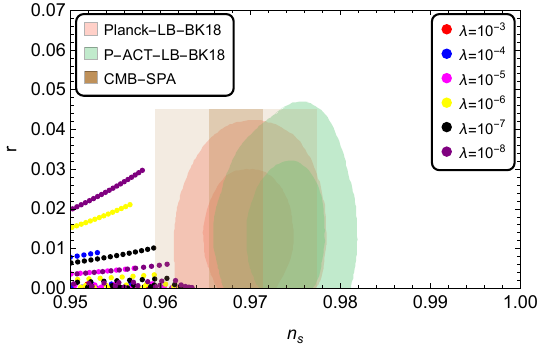}}
    \caption{The $r$--$n_s$ plane for the tanh-type Gauss-Bonnet coupling, with $\xi_1 \in [0, 0.03]$. Left: varying $n$. Right: varying $\lambda$. The background dark and light green contours show the P-ACT-LB-BK18 (Planck CMB anisotropies, ACT, combined with CMB lensing
[L], DESI DR2 BAO [B], and BICEP/Keck 2018 B-mode data [BK18]) $1\sigma$ and $2\sigma$ constraint regions. Whereas the dark and light pink shaded regions denote the $1\sigma$ and $2\sigma$ bounds of Planck-LB-BK18. The $1\sigma$ and $2\sigma$ bounds of the Planck-SPT-ACT (denoted as CMB-SPA) are represented by the dark and light brown rectangles, respectively~\cite{SPT-3G:2025bzu,Ellis:2025zrf}.} Unlike the exponential and sech cases, the tanh coupling struggles to place the model within the ACT $1\sigma$ region.
    \label{r_ns_tanh}
\end{figure*}

\begin{figure*}[t]
    \centering
    \subfigure[\label{nre_tre_exp}]{\includegraphics[width=0.45\linewidth]{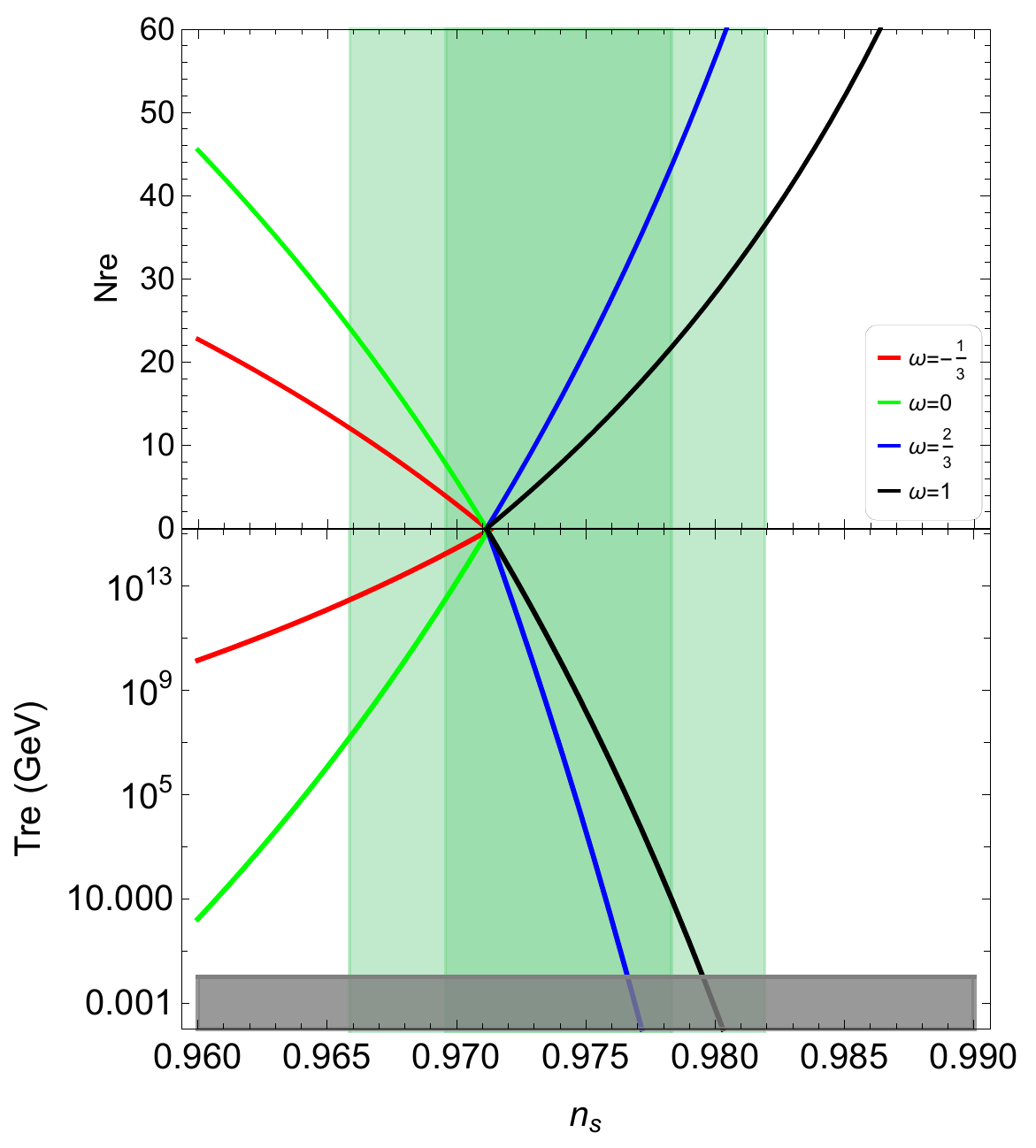}}
    \subfigure[\label{nre_tre_sech}]{\includegraphics[width=0.45\linewidth]{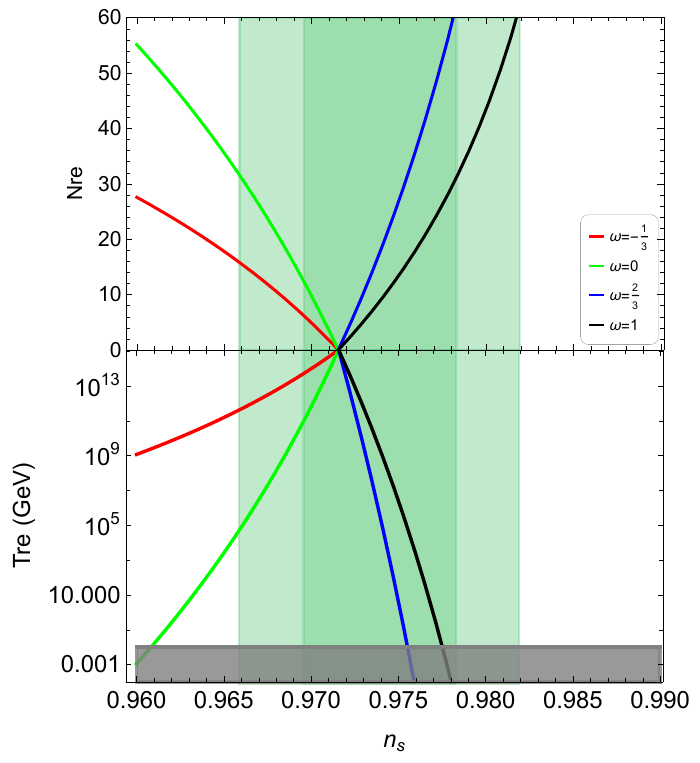}}
    \caption{Reheating parameters $N_{\rm re}$ vs.\ $T_{\rm re}$ for representative benchmark points. Left panel: exponential coupling with $n = 7$, $\lambda = 10^{-4}$, and $\xi_1 = 0.003$. Right panel: sech coupling with $n = 7$, $\lambda = 10^{-3}$, and $\xi_1 = 0.07$. In both cases, the reheating temperature satisfies the BBN lower bound. The shaded vertical dark and light green bands indicate the 1$-\sigma$ and 2$-\sigma$ ranges of ns from the P-ACT-LB-BK18 combination~\cite{ACT:2025fju,ACT:2025tim}.}
    \label{nre_tre_plot}
\end{figure*}

We now turn to the hyperbolic coupling functions. The analysis follows the same procedure as for the exponential case. For the sech coupling we vary $\xi_1 \in [0, 0.2]$, while for the tanh coupling we use $\xi_1 \in [0, 0.03]$; the potential parameters are kept in the same ranges as before ($n = 4$--$12$, $\lambda = 10^{-8}$--$10^{-3}$).

The results for the sech coupling are displayed in Fig.~\ref{r_ns_sech}. The predictions are well within the ACT $1\sigma$ region across a broad range of parameters, making this coupling function equally viable to the exponential form. By contrast, the tanh coupling (Fig.~\ref{r_ns_tanh}) consistently fails to bring the model into agreement with the ACT data for the parameter ranges explored. This stark difference between sech and tanh, despite both being smooth hyperbolic functions, underscores the importance of the functional form of the Gauss--Bonnet coupling in shaping the inflationary predictions.

The reheating analysis for the two viable coupling functions is shown in Fig.~\ref{nre_tre_plot}. For both the exponential and sech couplings, consistent reheating histories exist with $T_{\rm re}$ well above the BBN lower bound, confirming the internal consistency of the framework.

\section{Why the tanh coupling fails: an analytic argument}
\label{sec:tanh_failure}

The distinct behavior of the tanh coupling can be traced to a single 
algebraic sign, which then propagates consistently through every 
inflationary observable.

\paragraph{Sign of $\xi'(\phi)$.}
For the exponential and sech couplings in Eq.~\eqref{coupling_function},
the derivatives with respect to $\phi$ are
\begin{equation}
    \xi'_{\rm exp} = -\frac{\xi_1}{V_0}e^{-\xi_1\phi} < 0
    \end{equation}
    
    \begin{equation}
    \xi'_{\rm sech} = -\frac{\xi_1}{V_0}\,\mathrm{sech}(\xi_1\phi)
    \tanh(\xi_1\phi) < 0 \quad (\phi > 0),
\end{equation}
whereas for the tanh coupling
\begin{equation}
    \xi'_{\rm tanh} = \frac{\xi_1}{V_0}\,\mathrm{sech}^2(\xi_1\phi) > 0
    \quad \text{for all } \phi.
\end{equation}
This sign difference is the origin of the failure.

\paragraph{Propagation to $\delta_1$.}
The effective potential~\eqref{Veff} satisfies
\begin{equation}
    V'_{\rm eff} = \frac{U^2 V'}{V^2} + \frac{\xi'}{3}.
\end{equation}
For the quintessential potential $V = V_0 e^{-\lambda\phi^n}$ one has
$V' < 0$ throughout inflation. For the exponential and sech couplings,
both terms in $V'_{\rm eff}$ are negative, so $V'_{\rm eff} < 0$
robustly and inflation proceeds normally ($\epsilon_1 > 0$).
For the tanh coupling, the two terms compete in sign, but even in the 
regime where $V'_{\rm eff} < 0$ is maintained, the sign of $\delta_1$
is already reversed. From Eq.~\eqref{EGBsr3},
\begin{equation}
    \delta_1 = -\frac{2V^2}{3U^3}\xi' V'_{\rm eff}.
\end{equation}
With $V'_{\rm eff} < 0$: for exp/sech ($\xi' < 0$) one finds
$\delta_1 < 0$, while for tanh ($\xi' > 0$) one finds $\delta_1 > 0$.

\paragraph{Effect on $n_s$.}
The scalar spectral index~\eqref{nsVeff} can be written as
\begin{equation}
    n_s = 1 - 2\epsilon_1 - 
    \underbrace{\frac{2\epsilon_1\epsilon_2 - \delta_1\delta_2}
    {2\epsilon_1 - \delta_1}}_{\text{GB correction}}.
    \label{eq:ns_decomposed}
\end{equation}
In the standard-gravity limit ($\xi \to 0$, $\delta_1 \to 0$),
the quintessential potential gives $n_s^{\rm GR} \lesssim 0.965$, which lies more than $2-\sigma$ below the central value of the P-ACT-LB-Bk18 combination (though the tension is milder for Planck-alone or SPT-based dataset combinations. 
The GB term must therefore \emph{increase} $n_s$ to bring the prediction
into the $1\sigma$ region around $n_s^{\rm ACT} \simeq 0.974$.
This requires the GB correction in Eq.~\eqref{eq:ns_decomposed} to
be \emph{negative}.

For the exponential and sech couplings, $\delta_1 < 0$ and one can verify
that $\delta_2 < 0$ as well (since $V'_{\rm eff}$ 
and $\xi''$ both maintain the same sign). 
In the denominator, $\delta_1 < 0$ increases $|2\epsilon_1 - \delta_1|$,
so the correction is negative overall -- precisely the shift needed.

For the tanh coupling the situation is reversed: $\delta_1 > 0$
makes the correction positive, which \emph{decreases} $n_s$ rather than
increasing it. The Gauss--Bonnet term actively worsens the tension with 
ACT instead of resolving it, and no choice of $\xi_1$ in a
physically reasonable range can overcome this.

\paragraph{Asymptotic suppression of the tanh correction.}
A further obstacle is kinematic. Since $\xi'_{\rm tanh} = \xi_1\,
\mathrm{sech}^2(\xi_1\phi)$, the coupling derivative decays as
$e^{-2\xi_1\phi}$ at large field values. In a slow-roll trajectory
where $\phi$ is $\mathcal{O}(1)$ in Planck units at horizon crossing,
this suppression can be significant, effectively switching off the
Gauss--Bonnet contribution precisely at the scales probed by the CMB.
By contrast, the exponential coupling $e^{-\xi_1\phi}$ and the sech
coupling $\mathrm{sech}(\xi_1\phi)$ both decay more slowly and maintain
a non-negligible correction at the relevant field values.

\paragraph{Summary.}
The failure of the tanh coupling is not a numerical coincidence or a
consequence of an unlucky choice of parameters. It is a structural
consequence of the positive sign of $\xi'_{\rm tanh}$, which reverses
the sign of $\delta_1$ and thereby drives the Gauss-Bonnet correction
to $n_s$ in the wrong direction. Couplings that rescue quintessential
inflation from the P-ACT-LB-Bk18 tension must satisfy $\xi' < 0$ during inflation,
a condition met by the exponential and sech forms but violated by tanh.

\section{Conclusions}
\label{sec:conclusion}

In this work, we have revisited quintessential inflation in light of the precision measurements of the scalar spectral index from ACT DR6. 
The updated value $n_s = 0.9743 \pm 0.0034$ obtained from the combination of ACT DR6 with
Planck, CMB lensing, and DESI DR2 BAO and BK18 data (P-ACT-LB-Bk18) places the standard quintessential inflation model outside the $2\sigma$ allowed region, motivating an extension of the gravitational sector. We embedded the quintessential inflation potential in the EGB framework and studied the resulting inflationary dynamics in detail. The inclusion of a non-minimal scalar–Gauss–Bonnet coupling introduces new geometric degrees of freedom that substantially reshape the $r$--$n_s$ predictions. For suitable choices of the coupling function and free parameters, the EGB-modified model can be brought squarely within the $1\sigma$ region of the P-ACT-LB-BK18 constraints, a result that would be impossible within standard Einstein gravity. A systematic comparison of three coupling forms reveals a clear hierarchy: the exponential and sech-type couplings both provide viable parameter spaces in agreement with ACT, while the tanh coupling cannot accommodate the updated constraints. This sensitivity to the coupling structure is a generic feature of the EGB framework and suggests that observational data can be used to discriminate not just between different potentials, but between different geometric coupling functions (see Figs.~\ref{r_ns_exp},\ref{r_ns_sech} and Fig~\ref{r_ns_tanh})

We emphasize that our results depend on the particular combination of datasets used. Within the $\Lambda \mathrm{CDM}+r$ framework, the constituent datasets exhibit a mild preference for different values of $n_s$. In particular, a higher value of $n_s$ is obtained only when the Planck data are combined with ACT, DESI BAO, and BK18. By contrast, the latest SPT-3G measurements favor a value of $n_s$ that is much closer to the Planck. Consequently, if one considers only the Planck data or the Planck+SPT combination, quintessential inflation remains in good agreement with current observations. Our results should therefore be regarded as a proof of principle: if future observations establish the higher value of $n_s$ suggested by the ACT-based data combinations, Einstein-Gauss-Bonnet gravity offers a simple and economical mechanism for restoring the viability of quintessential inflation.
Beyond inflation, we examined the post-inflationary reheating phase in a model-independent manner. Even without a potential minimum, and hence without the usual oscillatory decay mechanism, both viable coupling forms admit consistent reheating histories with temperatures satisfying the BBN lower bound. This establishes the internal consistency of the full cosmological scenario from inflation through reheating. Taken together, our results indicate that EGB gravity offers a natural and robust pathway to reconcile quintessential inflation with the P-ACT-LB-BK18 dataset combinations. Rather than ruling out such potentials, the ACT results point toward the importance of the geometric sector -- and the EGB modification in particular -- in model building for the early universe. Looking ahead, improved upper bounds on the tensor-to-scalar ratio and better probes of the reheating epoch will provide decisive tests for this class of models, shedding further light on the interplay between high-energy physics and early-universe cosmology.
\acknowledgments
Yogesh would like to thank A. H. Mohammadi for useful discussions. MS and MRG are supported by the Science and Engineering Research Board (SERB), DST, Government of India, under the Grant Agreement number CRG/2022/004120 (Core Research Grant). MS is also partially supported by the Ministry of Education and Science of the Republic of Kazakhstan, Grant No. 0118RK00935. The authors would also like to thank the anonymous referee whose suggestions and comments have improved the overall quality of the paper.
\bibliographystyle{JHEP}
\bibliography{EGB_ACT}

\end{document}